\documentclass[prb,aps,final,twocolumn,floatfix,showpacs]{revtex4}
\usepackage{times}
\usepackage{pstricks}
\usepackage{graphicx}
\usepackage{gensymb}
\begin{document}

\title{Interplay between Raman shift and thermal expansion in
  graphene: temperature-dependent measurements and analysis of
  substrate corrections}

\author{S. Linas\footnote{These authors equally contributed to this
    work.}$^,$\footnote{Present address: LMI, Univ. Lyon 1 and CNRS
    UMR5615, 43 Bd du 11 Novembre 1918, 69622 Villeurbanne Cedex,
    France}}
\affiliation{ILM, Univ. Lyon 1 and CNRS UMR5306, 10 Rue Ada Byron,
  69622 Villeurbanne Cedex, France}
\author{Y. Magnin$^{*,}$\footnote{Present address: CINaM, CNRS and
    Aix-Marseille University, Campus de Luminy, Case 913, F13288
    Marseille, France}}
\affiliation{ILM, Univ. Lyon 1 and CNRS UMR5306, 10 Rue Ada Byron,
  69622 Villeurbanne Cedex, France}
\author{B. Poinsot}
\affiliation{ILM, Univ. Lyon 1 and CNRS UMR5306, 10 Rue Ada Byron,
  69622 Villeurbanne Cedex, France}
\author{O. Boisron}
\affiliation{ILM, Univ. Lyon 1 and CNRS UMR5306, 10 Rue Ada Byron,
  69622 Villeurbanne Cedex, France}
\author{G. D. F\"orster}
\affiliation{ILM, Univ. Lyon 1 and CNRS UMR5306, 10 Rue Ada Byron,
  69622 Villeurbanne Cedex, France}
\author{Z. Han}
\affiliation{Inst. N\'eel CNRS UPR2940 and Univ. Grenoble 1, Alpes,
  25 Rue des Martyrs, 38042 Grenoble, France}
\author{D. Kalita}
\affiliation{Inst. N\'eel CNRS UPR2940 and Univ. Grenoble 1, Alpes,
  25 Rue des Martyrs, 38042 Grenoble, France}
\author{V. Bouchiat}
\affiliation{Inst. N\'eel CNRS UPR2940 and Univ. Grenoble 1, Alpes,
  25 Rue des Martyrs, 38042 Grenoble, France}
\author{V. Martinez}
\affiliation{ILM, Univ. Lyon 1 and CNRS UMR5306, 10 Rue Ada Byron,
  69622 Villeurbanne Cedex, France}
\author{R. Fulcrand}
\affiliation{ILM, Univ. Lyon 1 and CNRS UMR5306, 10 Rue Ada Byron,
  69622 Villeurbanne Cedex, France}
\author{F. Tournus}
\affiliation{ILM, Univ. Lyon 1 and CNRS UMR5306, 10 Rue Ada Byron,
  69622 Villeurbanne Cedex, France}
\author{V. Dupuis}
\affiliation{ILM, Univ. Lyon 1 and CNRS UMR5306, 10 Rue Ada Byron,
  69622 Villeurbanne Cedex, France}
\author{F. Rabilloud}
\affiliation{ILM, Univ. Lyon 1 and CNRS UMR5306, 10 Rue Ada Byron,
  69622 Villeurbanne Cedex, France}
\author{L. Bardotti}
\affiliation{ILM, Univ. Lyon 1 and CNRS UMR5306, 10 Rue Ada Byron,
  69622 Villeurbanne Cedex, France}
\author{F. Calvo}
\affiliation{LIPhy, Univ. Grenoble 1 and CNRS, UMR5588, 140, Av. de la
  physique, 38402 St Martin D'H\`eres, France}

\begin{abstract}
Measurements and calculations have shown significant disagreement
regarding the sign and variations of the thermal expansion coefficient
(TEC) of graphene $\alpha(T)$. Here we report dedicated Raman
scattering experiments conducted for graphene monolayers deposited on
silicon nitride substrates and over the broad temperature range
150--900~K. The relation between those measurements for the G band and
the graphene TEC, which involves correcting the measured signal for
the mismatch contribution of the substrate, is analyzed based on
various theoretical candidates for $\alpha(T)$. Contrary to
calculations in the quasiharmonic approximation, a many-body potential
reparametrized for graphene correctly reproduces experimental
data. These results indicate that the TEC is more likely to be
positive above room temperature.
\end{abstract}

\maketitle

The thermal expansion coefficient (TEC) of materials involved in solid
interfaces is a key parameter characterizing the stress within the
materials, which in turn can modulate its electronic
properties.\cite{Ni08} The use of graphene in high density, integrated
electronic devices\cite{novoselov04,zhang05,bolotin08} or as matrix
reinforcement for composite materials\cite{stankovitch06} would
benefit from a better knowledge of the TEC, in particular its
dependence on temperature $\alpha(T)$.

Unfortunately, experiment and theory alike show markedly diverse
results regarding the TEC of graphene. The scanning electronic
microscopy (SEM) measurements carried by Bao {\em et al.} \cite{bao09}
found negative values for $\alpha$ at low temperature and a sign
change at around 350~K. Similar conclusions were reached by Yoon {\em
  et al.}\cite{yoon11} who performed Raman scattering spectroscopy,
although no sign change was observed below 400~K. Negative
coefficients were also obtained by Singh {\em et al.} \cite{singh10}
who used a nanoelectromechanical resonator. In both
Refs.~\onlinecite{yoon11} and \onlinecite{singh10} strong variations
among samples were emphasized. On the theory side, density-functional
perturbation theory (DFPT)\cite{mounet05} and ab initio molecular
dynamics \cite{pozzo11} predicted an all negative $\alpha$ in a broad
temperature range, whereas non-equilibrium Green's function
calculations\cite{jiang09} found a sign change near 600~K. Atomistic
Monte Carlo (MC) simulations have found all positive, all negative or sign
changing variations of the TEC depending on the potential
used.\cite{zakharchenko11,magnin14} This diversity of behaviors is
related to the importance of anharmonicities\cite{fasolino07} and the
difficulty of describing them properly in relation with appropriate
measurements.

Raman spectroscopy in a broad temperature range is one of the indirect
ways to access such properties. Being a fast and non destructive tool
that can offer structural and electronic informations, it has been
widely used for the characterization of graphene.\cite{ferrari13} In
particular, from such measurements the number of layers, density of
defects, amount of stress and doping can all be
evaluated.\cite{ferrari13} As a direct probe of the phonon structure,
temperature-dependent measurements should be indirectly related to the
lattice parameter, hence to the TEC. However, in experiments graphene
is held on (or by) a substrate, and it is well known that the detailed
graphene structure is sensitive to the nature of the
support.\cite{soldano10} In particular, incommensurability between the
two lattices gives rise to strain often manifested by
corrugation,\cite{Preobrajenski08} and which could affect the measured
TEC.\cite{jean13} More generally, the contact between the two
materials having different thermal expansion coefficients is a source
of strain.\cite{yoon11} Even though some authors have disregarded this
correction in their measurements,\cite{bao09} the importance of
substrate interactions on the TEC has been recognized
before.\cite{jiang09}

One limitation of earlier investigations is the rather restricted
temperature ranges over which the measurements were conducted,
generally below 400~K. In the present work, we have extended this
range to an higher upper limit of 900~K. More importantly, we have
carried out a comprehensive analysis of the Raman G band based on
underlying models for the graphene TEC, carefully disentangling the
contribution of the substrate by following the phenomenological
procedure laid out by Yoon {\em et al.}\cite{yoon11} Our experimental
results are found incompatible with TEC 
that remain negative in the entire temperature range, but agree
reasonably well with an all-positive model TEC predicted by a
dedicated atomistic potential precisely fitted to reproduce the phonon
structure of graphene.

Strictly monolayer graphene was synthesized on a copper foil (25
$\mathrm{\mu}$m thick, 99.8\%~purity, AlfaAesar) by a pulsed CVD
growth method.\cite{han13} After etching of the foil in a
$\mathrm{(NH_{4})_{2}S_{2}O_{8}}$ solution at $5 \times
10^{-2}$~mol/L concentration, the graphene sheet was directly
transferred on a SiN membrane (thickness 50~nm) supported on a silicon
substrate (Silson) using a resist-free technique.\cite{kim09} Before
experiment, the samples were annealed \emph{in-situ} at 600~K in an
inert atmosphere. The measurements were carried out using two distinct
Raman setups. First, a Renishaw RM~1000 micro-Raman spectrometer
equipped with a 1800~lines/mm grating used a laser power kept low
enough not to induce any shift of the G peak. A $\times$50 long
working distance objective was used, and the samples were heated and
cooled in a Linkam THMSG600 in a ultra pure Ar (alphagaz~2, Air
Liquide) atmosphere. A second homemade Raman setup was adapted to a
ultrahigh vacuum (UHV) chamber (base pressure $10^{-9}$~mbar) and a
Horiba spectrometer (TRIAX320). The excitation wavelength was set to
532~nm in both experiments. The G and 2D peaks were fitted to single
Lorentzians for analysis. Additional SEM imaging was performed after
the Raman measurements in a FEI, NovaSEM 450 microscope at an
acceleration voltage of 5~kV and with a sample tilted at 45{\degree}
with respect to the electron beam.

\begin{figure}[htb]
  \centerline{\includegraphics[width=8.5cm]{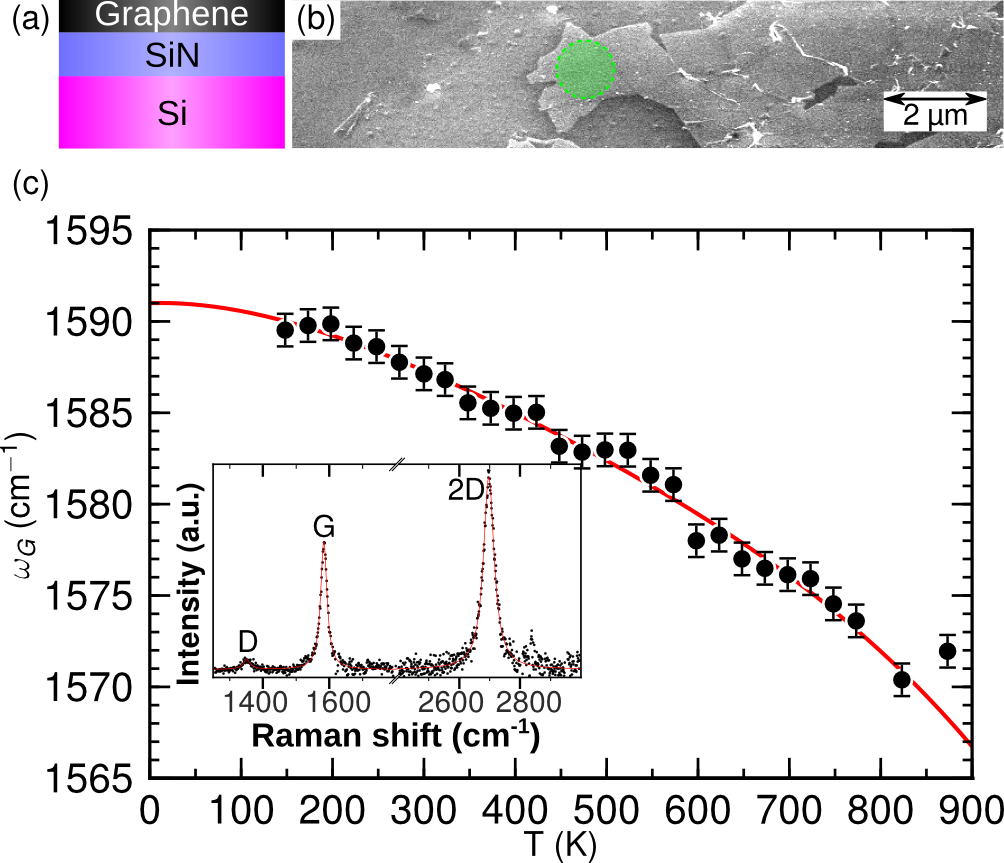}}
  \caption{(Color online) (a) Schematic of the sample, a single layer
    graphene film grown by CVD and transferred on a SiN/Si
    substrate. (b) Scanning electron micrograph of the sample, the
    probed area being highlighted by a black dashed circle. (c) G peak
    frequency ($\omega_G$, symbols) fitted to 0~K using a 
    polynomial fit (red solid line). The inset shows a typical Raman
    spectrum at $T=300$~K (black circles) and Lorentzians fits of the
    D, G and 2D peaks (red line).}
  \label{fig:fig1}
\end{figure}

In Fig.~\ref{fig:fig1}, panel (a) schematically depicts the graphene
sample supported on the SiN/Si substrate, while panel (b) shows a SEM
micrograph including the Raman laser spot. A typical Raman spectrum
represented in the inset of Fig.~\ref{fig:fig1}(c) shows symmetric
Lorentzian line shapes for the 2D peak and an intensity ratio
$I_{2D}/I_{G}\sim 1.5$ with the G peak which are consistent with a single
graphene layer.\cite{ferrari13} In addition, the low ratio $I_D/I_G$
with the D peak suggests a low density of defects.\cite{ferrari13}

The variations of the G peak frequency $\omega_G(T)$ with increasing
temperature are shown as the main Fig.~\ref{fig:fig1}(c) for a sample
in argon atmosphere. These reasonably smooth variations, together with
the SEM data, indicate that the graphene layer does not present
significant folded areas, cracks or wrinkles at least in the area
probed by the laser. Additional measurements performed on other
samples and under UHV or argon atmosphere show comparable results with
a good reproducibility.\cite{SI}

The room temperature value of $\omega_G$ found for our sample
(1587.2~cm$^{-1}$) is slightly shifted compared to the intrinsic value
of 1581.6~cm$^{-1}$ expected for charge- and strain-free
graphene,\cite{lee12} suggesting mild amounts of defects in our
sample. The small discontinuities of 1--2~cm$^{-1}$ observed for
$\omega_G(T)$ might originate from a stick-slip of the graphene layer
on the nitride surface, as also visible on the measurements of
Calizo and coworkers on silica substrates.\cite{calizo07} Between 100
and 400~K the temperature variations of $\omega_G$ are roughly linear
with a slope of $-0.023~$cm$^{-1}$.K$^{-1}$, in fair agreement with
values reported previously for graphene deposited on silica substrates
($-0.016$~cm$^{-1}$.K$^{-1}$ in Ref. \onlinecite{calizo07},
$-0.05$~cm$^{-1}$.K$^{-1}$ in Ref. \onlinecite{yoon11}).

The observed thermal contribution to the Raman frequency shift of the
G peak, $\Delta\omega_G(T)$, was evaluated by removing from
$\omega_G(T)$ the value extrapolated at 0~K using a polynomial fit,
$\omega_G(T=T_0\simeq 0)=1591$~cm$^{-1}$ (note that this fit serves no
other purposes than extrapolating to the $T_0$ reference
temperature). To determine the intrinsic Raman shift of the pure
graphene layer, the contribution $\Delta \omega^S_G(T)$ of the
substrate-induced strain was removed from $\Delta \omega_G(T)$
following the same procedure advocated by Yoon {\em et
  al.}\cite{yoon11} based on the TECs of both graphene and substrate,
\begin{equation}
\Delta \omega^{S}_G(T)=\beta \varepsilon(T) = \beta
\int^T_{T_0} \left[ \alpha_{\rm sub}(T) - \alpha_{\rm gr}(T) \right]
dT.
\label{eq:strain}
\end{equation}
In the previous equation we have denoted by $\alpha_{\rm sub}$ and
$\alpha_{\rm gr}$ the TECs of the substrate and of the graphene layer,
respectively. $\beta$ is the biaxial strain coefficient of the G band
known to be approximately\cite{mohiuddin09,yoon11_2} $\beta =-70 \pm
3$~cm$^{-1}$/\%. Eq.~(\ref{eq:strain}) is central to our analysis
as it provides a relation between the experimentally measured Raman
signal and the graphene TEC we aim to discuss. To evaluate the
substrate contribution to the Raman signal it is necessary to
integrate the thermal expansion coefficients of the two materials over
temperature.  The TEC $\alpha_{\rm sub}$ of the SiN substrate was
taken from the literature\cite{sinha78} and extrapolated down to low
temperatures $T<400$~K using data known for the similar Si$_3$N$_4$
system.\cite{si3n4}

\begin{figure}[htb]
  \centerline{\includegraphics[width=8.5cm]{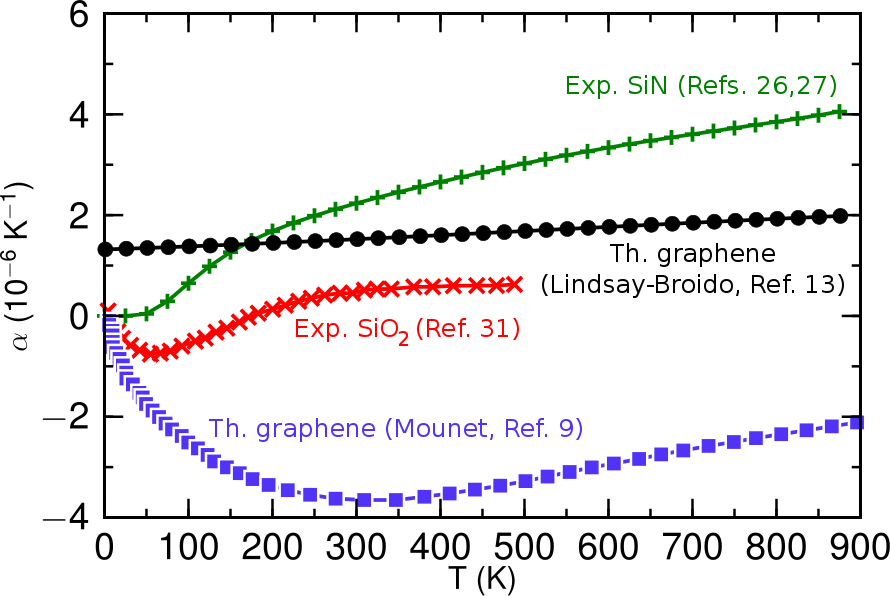}}
  \caption{(Color online) Temperature dependence of the experimental
    thermal expansion coefficients $\alpha$ of silicon nitride (green
    pluses) and silica (red crosses), and theoretical in-plane
    coefficients of graphene obtained by Mounet and Marzari in the
    quasiharmonic approximation (blue squares) and from classical
    MC simulations based on the Lindsay-Broido
    reparametrization of the Tersoff bond-order potential for graphene
    (black circles).}
  \label{fig:CTE}
\end{figure}

\begin{figure}[htb]
  \centerline{\includegraphics[width=8.5cm]{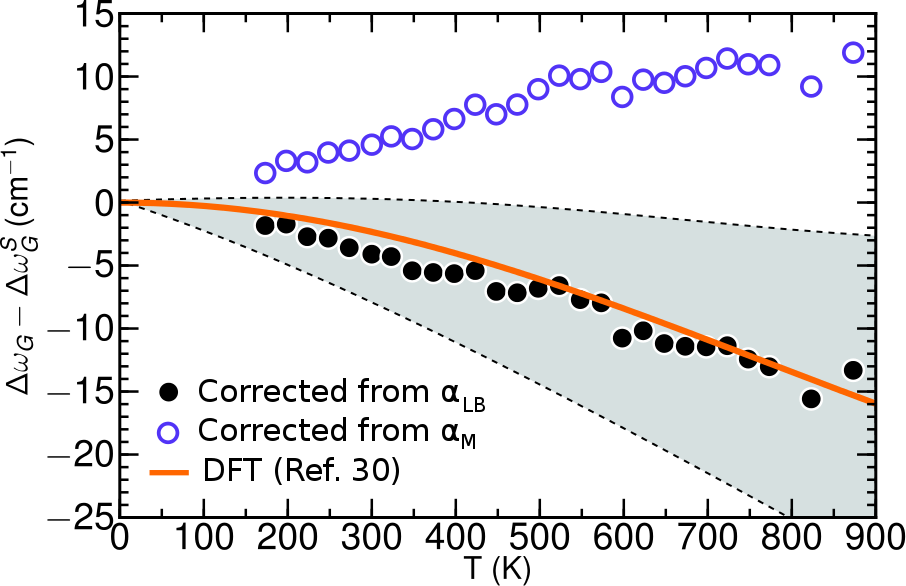}}
  \caption{(Color online) Temperature dependence of the Raman G band
    shift of pure graphene, corrected from the substrate mismatch
    contribution using different model TEC $\alpha_{\rm gr}$, namely
    the quasiharmonic Mounet-Marzari model ($\alpha_{\rm M}$, empty
    circles) and the Lindsay-Broido semiempirical model ($\alpha_{\rm
      LB}$, full circles). The shaded area corresponds to varying the
    $\alpha_{\rm LB}$ function by $\pm 2\times 10^{-6}$~K$^{-1}$, and
    the solid line shows the Raman shift calculated by Bonini and
    coworkers\cite{bonini07} from DFT.}
  \label{fig:dwlocal}
\end{figure}

For graphene, several forms were tried for $\alpha_{\rm gr}(T)$ in the
hope that comparison with experiment would ultimately settle generic
conclusions about the expected features of this fundamental quantity.
The DFPT results from Mounet and Marzari\cite{mounet05} were chosen as
a representative of the quasiharmonic approximation based on
first-principles data, giving a TEC that we denote as $\alpha_{\rm M}$
and that is entirely negative in the relevant temperature
range. Alternatively, among the various predictions of fully
anharmonic MC simulations based on atomistic
potentials\cite{magnin14} we have chosen those obtained with a recent
reparametrization of the Tersoff bond-order potential\cite{tersoff} by
Lindsay and Broido\cite{lindsay10} dedicated to graphene, $\alpha_{\rm
  LB}$. The variations of the three aforementioned thermal expansion
coefficients with temperature are represented in Fig.~\ref{fig:CTE}.
The strong discrepancies between the two model TECs for graphene are
expected to convey to the Raman shift, and we have reported in
Fig.~\ref{fig:dwlocal} the variations of $\Delta \omega_G(T)$ obtained
by integrating Eq.~(\ref{eq:strain}) with the corresponding functions
$\alpha_{\rm gr}=\alpha_{\rm M}$ or $\alpha_{\rm LB}$.
In order to compare the relative performances of the two models, a
third set of reference data is required, which is provided by the
theoretical Raman shift of freestanding graphene calculated by Bonini
and coworkers\cite{bonini07} using density-functional theory (DFT)
calculations under appropriate anharmonic expansions. This purely
theoretical result, depicted also in Fig.~\ref{fig:dwlocal} clearly
agrees quantitatively with the present measurements if the substrate
correction originates from the Lindsay-Broido model, but disagrees
otherwise. It is important to evaluate the sensitivity of this result
to the $\alpha_{\rm gr}$ ingredient, and we have repeated the
integration using the $\alpha_{\rm LB}$ model but shifting it by $\pm
2\times 10^{-6}$~K$^{-1}$ in the entire temperature range, a negative
shift leading to a positive TEC at low temperature and a sign change
near 400~K. The resulting Raman shift of the G band also varies (see
Fig.~\ref{fig:dwlocal}) but remains closer to the reference DFT data
than the values obtained with the $\alpha_{\rm M}$ correction. This
observation puts some constraints on the true $\alpha_{\rm gr}$
function.

\begin{figure}[htb]
  \centerline{\includegraphics[width=8.5cm]{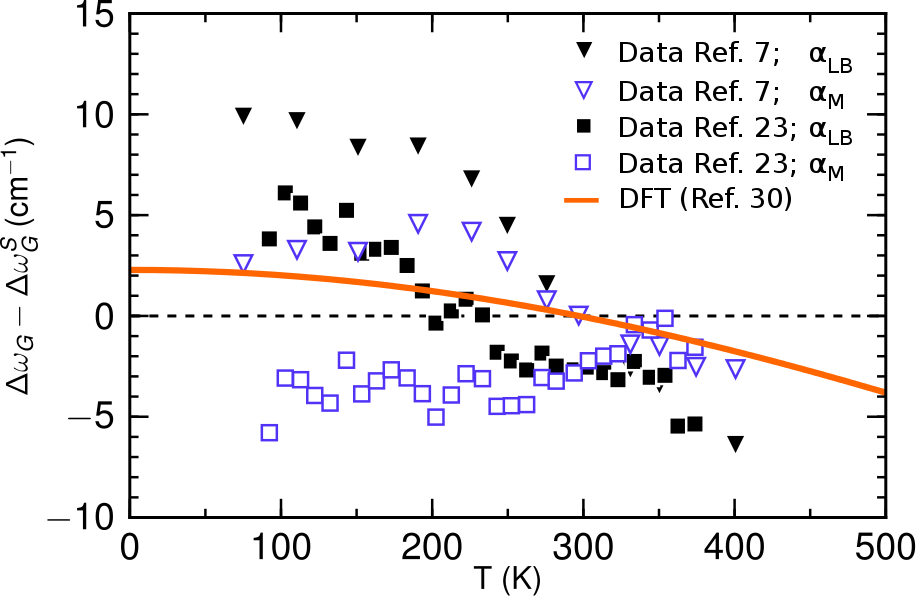}}
  \caption{(Color online) Temperature dependence of the Raman G band
    shift of pure graphene from measurements on the
    SiO$_2$ substrate, as measured by Yoon {\em et al.}
    (Ref. \onlinecite{yoon11}, triangles pointing down) and by Calizo
    {\em et al.} (Ref. \onlinecite{calizo07}, squares) and corrected
    for the substrate mismatch contribution using the Mounet-Marzari
    ($\alpha_{\rm M}$, empty symbols) and Lindsay-Broido ($\alpha_{\rm
      LB}$, full symbols) models. The solid line shows the Raman shift
    calculated by Bonini and coworkers\cite{bonini07} from DFT
    calculations, shifted to cross the vertical axis at 297~K.}
  \label{fig:dwexternal}
\end{figure}

According to the present measurements and analysis, the graphene TEC
is better described by the Lindsay-Broido model than the
Mounet-Marzari quasiharmonic model. It is tempting to challenge those
conclusions by considering the case of silica substrates, on which
earlier Raman measurements have been performed.\cite{yoon11,calizo07}
Two sets of experimental data were borrowed from the works of Yoon and
coworkers\cite{yoon11} and Calizo {\em et al.},\cite{calizo07}
respectively, and subject to the same correcting treatment as
performed here for silicon nitride but using the experimental thermal
expansion coefficient of silica\cite{sio2} also superimposed in
Fig.~\ref{fig:CTE}. From these two data sets for $\Delta\omega_G(T)$,
the correction $\Delta \omega_G^S(T)$ was calculated from the two
graphene model TECs $\alpha_{\rm M}$ and $\alpha_{\rm LB}$, and the
results are again compared to the reference DFT results of Bonini and
coworkers\cite{bonini07} in Fig.~\ref{fig:dwexternal}.
Correcting the data by Yoon {\em et al.} with the Mounet-Marzari model
for $\alpha_{\rm gr}$ leads to values for the Raman shift that are in
good agreement with those obtained in Ref.~\onlinecite{yoon11}, as
expected. They are significantly closer to the DFT reference data than
those obtained with the all-positive TEC predicted by the
Lindsay-Broido model, although the agreement is never
excellent. Incidentally we note that Yoon and coworkers had to
significantly adjust the input $\alpha_{\rm gr}(T)$ in order to get an
even better agreement. However, the opposite observations can be made
when the calculations are performed using the experimental data from
Calizo {\em et al.},\cite{calizo07} and a fully negative model for
$\alpha_{\rm gr}$ markedly underestimates the Raman shift while the
Lindsay-Broido model yields reasonable agreement. Given the
significant dispersion among experimental measurements for graphene on
SiO$_2$ substrates,\cite{yoon11,calizo07} and in particular the much
more limited temperature range on which these measurements have been
conducted, it seems difficult to conclude unambiguously based on the
silica data alone whereas the present data obtained on silicon nitride
display better overall agreement with theoretical models.

The discrepancies between the predictions of the Mounet-Marzari and
Lindsay-Broido corrections are related to the rather different nature
of the two models, the former quasiharmonic model\cite{mounet05} being
expected to be better at low temperature while the latter, anharmonic
but classical model\cite{lindsay10} was adjusted on room temperature
properties. Both models were not fitted to reproduce
temperature-dependent properties, hence it is also unclear to which
extent they would be able to correctly capture anharmonicities
especially above room temperature. In absence of independent
measurements, and in view of the major diversity among computational
models,\cite{magnin14} it is tempting to conclude that the true
thermal expansion coefficient lies somewhere inbetween the two models,
hence that it must at least be positive above some temperature close
to 300~K.

In summary, we have performed Raman scattering measurements of
graphene monolayers supported on silicon nitride over an extended
temperature range, and used these data to establish some constraints
on the TEC of pure graphene. Our analysis relies on the correction to
the measured Raman shift of the G band due to the mismatch
contribution from the substrate, and on the comparison of the
corrected signal to benchmark anharmonic DFT
calculations.\cite{bonini07} The present approach, which follows
earlier efforts by other authors,\cite{yoon11} requires knowledge of
the expansion coefficients of both graphene and substrate materials. A
model TEC based on finite temperature MC simulations with a potential
dedicated to graphene turned out to reproduce best the experimental
data, suggesting that the TEC is more likely to be positive above
moderate temperatures. Until more direct measurements are carried out,
one main outcome of the present work is the confirmation that the
substrate plays a great role on the thermal properties of
graphene,\cite{jiang09} and this contribution should not be neglected
in general especially when attempting some theoretical
predictions. Future work could be devoted to extending the present
methodology to other 2D or layered materials such as hexagonal boron
nitride or to transition metal dichalcogenides that currently hold
promises for their interesting semiconductor properties.\cite{wang12}

The authors gratefully acknowledge the CECOMO (Univ. Lyon 1) for
providing access to the Raman spectroscopy facilities and C. Albin
(ILM, PLYRA). Research supported by ANR Contract
No. ANR-2010-BLAN-1019-NMGEM.

\end{document}